\documentclass{iaus}
\usepackage{graphicx}

\title[S282~~Binaries in Other Galaxies] 
{Techniques for Observing Binaries in Other Galaxies}

\author[Alceste Z. Bonanos]   
{Alceste Z. Bonanos$^1$}

\affiliation{$^1$National Observatory of Athens, Institute of Astronomy
  \& Astrophysics, \\ I. Metaxa \& Vas. Pavlou, Palaia Penteli 15236, Greece \\ email: {\tt bonanos@astro.noa.gr}}

\pubyear{2011}
\volume{282}  
\pagerange{119--126}
\jname{From Interacting Binaries to Exoplanets: Essential Modeling Tools}
\editors{A.C. Editor, B.D. Editor \& C.E. Editor, eds.}
\begin{document}

\maketitle

\begin{abstract}

I present an overview of the techniques used for detecting and following
up binaries in nearby galaxies and present the current census of
extragalactic binaries, with a focus on eclipsing systems. The
motivation for looking in other galaxies is the use of eclipsing
binaries as distance indicators and as probes of the most massive stars.


\end{abstract}

\firstsection 
\section{Motivation}

Eclipsing binaries are not only powerful tools for obtaining fundamental
parameters of stars (\cite[Andersen 1991]{Andersen91}, \cite[Torres et
  al. 2010]{Torres10}), but also the most accurate tools currently
available for measuring masses and radii of massive stars and for
probing the upper stellar mass limit. Double-lined spectroscopic binary
systems exhibiting eclipses in their light curves provide accurate
geometric measurements of the fundamental parameters of their component
stars. Specifically, the light curve provides the orbital period,
inclination, eccentricity, the fractional radii and flux ratio of the
two stars. The radial velocity semi-amplitudes determine the mass ratio;
the individual masses can be solved using Kepler's third
law. Furthermore, by fitting synthetic spectra to the observed ones, one
can infer the effective temperatures of the stars, solve for their
luminosities and derive the distance (e.g. \cite[Bonanos
  \etal\ 2006]{Bonanos06}). In the past two decades, many eclipsing
binaries have been discovered in other galaxies and several of these
have been subject to follow up studies, resulting in the measurement of
their fundamental parameters. The main motivations for observing
eclipsing binaries in other galaxies are to study massive stars and to
obtain independent distances.

Massive stars are intrinsically rare compared to their lower mass
counterparts, due to their shorter lifetimes and the steep initial mass
function, which results in the formation of a smaller number of massive
stars. Studying massive stars in the Galaxy is challenging, because they
are located in the Galactic plane, where they reside in young massive
clusters and usually near giant molecular clouds, and are therefore
often heavily obscured by dust. Fig. 10 of \cite{Mauerhan11}
demonstrates the small fraction of the Milky Way surveyed for massive
stars, by showing the locations of known Wolf-Rayet (WR) stars in the
Galaxy. Although the total estimated number of WR stars in the Galaxy is
6500, there are only $\sim600$ known and most are located within 5 kpc
of the Sun, i.e. only $\sim10\%$ of the Milky Way has been
surveyed. This fraction is slowly increasing, with the recent
availability of near-infrared and mid-infrared maps of the Galactic
plane (obtained with {\it Spitzer}), which have been used both to
identify new massive clusters (e.g. \cite{Davies07}) and massive evolved
stars with nebulae (\cite{Gvaramadze10}, \cite{Wachter10}).

Another reason why our knowledge of massive stars is incomplete is
because their fundamental parameters are not well known, leaving
formation and evolution models unconstrained. Masses and radii of
massive stars measured from eclipsing binaries remain
scarce. \cite{Bonanos09} compiled a list of the most massive stars
accurately measured in eclipsing binary systems and found only 14 stars
above 30 Mo with mass and radius measurements accurate to 10\% or
better. Since this compilation, measurements of only 2 more massive
stars (\cite{Stroud10}) satisfy these accuracy requirements, bringing
the total to 16. Therefore the need for accurate fundamental parameters
of very massive stars at a range of metallicities and evolutionary
phases remains of primary importance.

There are several advantages to studying massive stars in other
galaxies, despite their greater distance from us. The low foreground
extinction allows observations in the optical and ultraviolet, where the
stars emit the most light, making possible their identification and
study with smaller telescopes. The large metallicity range found in
Local Group galaxies and beyond allows for a comparative study of the
properties of massive stars as a function of metallicity (see
e.g. \cite{Massey03}), which is an important factor determining their
fate. As variability studies become more widespread, eclipsing binaries
are being identified in an increasing number of galaxies. Follow up
studies of massive extragalactic systems is crucial to our understanding
of massive star evolution. 

Studying massive stars in other galaxies also offers the opportunity to
obtain a complete census of eclipsing binaries and statistics on the
binarity of whole populations of massive stars, a task that is currently
impossible in our Galaxy. Specifically, obtaining the complete number of
eclipsing systems in a galaxy down to a certain magnitude and within a
certain period range will help constrain the binarity fraction of the
higher mass population, which is near 50\% among massive stars
(\cite{Sana10}).

Finally, another motivation for studying eclipsing binaries in other
galaxies is that they are good distance indicators (\cite{Paczynski97}),
which can provide independent and accurate distances to Local Group
galaxies. Given the radius and effective temperature of the component
stars of the system, their luminosities (or absolute magnitude) can be
calculated. Armed with both the absolute and apparent magnitude, and
after correcting for extinction, one can obtain the distance.

\section{Techniques}

The most efficient techniques for observing eclipsing binaries in other
galaxies, while not vastly different from galactic studies, include
photometric variability studies with wide field CCDs and follow-up
observations with multi-object spectrographs. Difference imaging or
image subtraction (\cite{Alard98}, \cite{Alard00}) is a technique widely
used in extragalactic variability studies, given the crowded nature of
the fields. \cite{Bonanos03} demonstrated it to be a much more efficient
method for detecting variables in crowded fields compared with
traditional PSF-fitting photometry.

The discovery of extragalactic eclipsing binaries mainly comes from
variability surveys of nearby galaxies with 1-2 meter telescopes, such
as the DIRECT project (\cite{Stanek98}, \cite[Bonanos
  \etal\ 2003]{Bonanos03a}) that specifically aimed to discover
eclipsing binaries in M31 and M33, or the Araucaria Project
(e.g. \cite{Pietrzynski02}), which is surveying several nearby galaxies
for RR Lyrae, Cepheids and eclipsing binaries to obtain accurate
distances. Large numbers of extragalactic binaries have also resulted as
side products of microlensing surveys, such as MACHO and OGLE, which
have discovered thousands of eclipsing binaries in the Magellanic Clouds
(see \cite{Derekas07} and \cite{Faccioli07} for MACHO results, and
\cite{Wyrzykowski03}, \cite{Wyrzykowski04} for OGLE-II
results). Furthermore, the long time baseline of the OGLE project has
resulted in the discovery of very long period systems or other rare
systems, such as an eclipsing system containing a Cepheid
(\cite{Pietrzynski10}).

Once the eclipsing binaries have been identified via photometric
variability surveys, 6-10 meter class telescopes are needed for
follow-up spectroscopic observations. Service mode observing (available
e.g. at Gemini, VLT), targeting quadrature phases has been shown to be
the most efficient way of obtaining spectroscopy for a small number of
targets per galaxy (\cite{Gonzalez05}). When the number of targets is
large (e.g. \cite[Hilditch \etal\ 2005]{Hilditch05}), then multi-object
spectrographs, such as FLAMES/VLT or 2dF/AAT, provide the most efficient
follow up method. With the currently available telescopes, fundamental
parameters of eclipsing binaries can be measured out to a distance limit
of about 1 Mpc, as a resolving power $R\geq3000$ and S/N$\geq30$ are
necessary for early-type systems and targets typically have $V>18$ mag.

Last but not least, several multi-epoch spectroscopic surveys have been
undertaken to identify spectroscopic binaries (e.g. \cite{Foellmi03}),
some of which are later found to be eclipsing systems as well
(e.g. WR20a, \cite{Rauw04}, \cite{Bonanos04}). The VLT-FLAMES Tarantula
survey (\cite{Evans11}) is a recent example of such a multi-epoch
spectroscopic survey, with the goal to identify massive binaries via
radial velocity variations.

\section{Eclipsing Binaries in Other Galaxies}

Table 1 presents a census of known extragalactic eclipsing binaries. The
first six galaxies are Local Group members, while NGC 300 is in the
Sculptor group and NGC 2403 in the M81 group. The eclipsing binary in
NGC 2403 was discovered by \cite{Tammann68} and has a B magnitude of 22.

\begin{table}[ht]
  \begin{center}
  \caption{Census of Extragalactic Eclipsing Binaries.}
  \label{tab1}
 {\small
  \begin{tabular}{|l|c|c|c|}\hline 
{\bf Galaxy} & {\bf Distance} & {\bf \# of EBs} & {\bf Source}  \\ 
\hline
LMC & 50 kpc & 4634, 2580 & MACHO, OGLE \\ \hline
SMC & 60 kpc & 1509, 1350 & MACHO, OGLE \\ \hline
NGC 6822 & 460 kpc& 3 & Araucaria Project \\ \hline
IC 1613 & 730 kpc& 1 & Araucaria Project  \\ \hline
M31 & 750 kpc& $\sim500$ & DIRECT Project \& \cite{Ribas04} \\ \hline
M33 & 960 kpc& 148 & DIRECT Project  \\ \hline
NGC 300 & 1.9 Mpc & 1 & Araucaria Project \\ \hline
NGC 2403 & 2.5 Mpc & 1 & \cite{Tammann68} \\ \hline
  \end{tabular}
  }
 \end{center}
\vspace{1mm}
\end{table}

The large number of systems in the Magellanic Clouds is due to the MACHO
and OGLE microlensing surveys. \cite[Faccioli \etal\ (2007)]{Faccioli07}
presented a catalog of MACHO eclipsing binaries, while \cite[Wyrzykowski
  \etal\ (2003)]{Wyrzykowski03} and \cite[Wyrzykowski
  \etal\ (2004)]{Wyrzykowski04} presented the catalogs from the OGLE
survey. While some of these are bound to be foreground systems, most are
indeed extragalactic. Note, there is some overlap between the
catalogs. Moving farther out, the dwarf galaxy eclipsing systems in IC
1613 and NGC 6822 were discovered by the Araucaria project. Finally, the
significant number of systems discovered in M31 and M33 is due to the
dedicated searches by the DIRECT Project (e.g. \cite{Stanek98}) and
\cite{Ribas04}.

The Local Group eclipsing binaries lend themselves as distance
indicators and have been used as such so far to derive distances to the
LMC, SMC, M31 and M33. \cite{Guinan98}, \cite{Ribas02},
\cite{Fitzpatrick02}, \cite{Fitzpatrick03} have used early-B type
systems to derive eclipsing binary distances to the LMC, while
\cite{Pietrzynski09} used a G-giant eclipsing system and
\cite{Bonanos11} an O-type eclipsing system. Most systems in the bar of
the LMC are found to be at 50 kpc, however the distance to HV 5936 is
discrepant, likely due to the 3-dimensional structure of the galaxy. In
the SMC, \cite{Harries03} and \cite{Hilditch05} have obtained a distance
modulus of 18.91 $\pm$ 0.03 mag by measuring 50 OGLE-II binaries with
AAT/2dF spectrograph, while \cite{North10} obtained a distance modulus
of 19.11 $\pm$ 0.03 mag with 33 OGLE-II eclipsing binaries, using
VLT/FLAMES. The discrepancy in the distance likely arises from
systematic errors associated with lower resolution spectra from 2dF and
the estimation of the extinction. 

In M31, the eclipsing binary distances of \cite[Ribas \etal\ (2005, 772
  $\pm$ 44 kpc or 24.44 $\pm$ 0.12 mag)]{Ribas05} and \cite[Vilardell
  \etal\ (2010, 724 $\pm$ 37 kpc or 24.30 $\pm$ 0.11 mag)]{Vilardell10}
are in agreement with each other. However, in M33, the long distance
derived by \cite{Bonanos06}, 960 $\pm$ 54 kpc, was not in agreement with
most measurements in the literature, and in particular with the HST Key
Project measurement (\cite{Freedman01}), possibly because of the
difficulty in estimating reddening with other methods. Nonetheless, the
M33 result has pushed our current capabilities to the limit, measuring
fundamental parameters of stars out to 1 Mpc. Overall, eclipsing binary
distances are very valuable, because they provide independent distances,
which can help evaluate the systematic errors associated with other
widely used standard candles (e.g. Cepheids, RR Lyrae, tip of the red
giant branch).

\section{Future}

The potential of eclipsing binaries for obtaining fundamental parameters
of massive stars and independent distances to other galaxies is
extremely promising. The ongoing OGLE project, now in its phase IV, is
surveying even larger areas of the Magellanic Clouds and is bound to
discover tens of thousands of eclipsing binaries. Furthermore, transient
surveys such as Pan-STARRS and the Palomar Transient Factory, as well as
asteroid surveys, such as the Catalina Sky Survey, and in the future,
the Large Synoptic Sky Telescope will be including many nearby galaxies
in their fields and monitoring them for long periods of time.

In conclusion, wide field surveys and multi object spectrographs are
truly revolutionizing extragalactic binary studies. The rate of
discovery of such systems is bound to increase and provide ample
opportunity for studies of extragalactic massive stars, the
determination of their distances, the binarity fraction and finally,
statistics on binarity of various populations of stars in nearby
galaxies.

\begin{discussion}

\vspace{0.3cm}

\noindent R. E. WILSON: The way you find distances to eclipsing binaries
is very good and logical (using complete optical light curves for most
parameters and then the few infrared points for distance, thereby being
relatively free of interstellar extinction dependence). However, now one
can go a bit further, as the 2010 version of the WD program avoids the
spherical star approximation previously used with the infrared points in
the distance step. The program also gives options (process the optical
and infrared data separately or together, or both ways) and assumes
consistency. It is directly absolute, with fluxes in physical units and,
since the program does most of the work, it makes the overall process
very fast.

\end{discussion}

\end{document}